# Title: Observation of Chern insulator in crystalline ABCA-tetralayer graphene with spin-orbit coupling

**Authors:** Yating Sha[1*], Jian Zheng[1*], Kai Liu[1*], Hong Du[2], Kenji Watanabe[3], Takashi Taniguchi[4], Jinfeng Jia[1], Zhiwen Shi[1], Ruidan Zhong[2], Guorui Chen[1]†

[1]Key Laboratory of Artificial Structures and Quantum Control (Ministry of Education), Shenyang National Laboratory for Materials Science; School of Physics and Astronomy, Shanghai Jiao Tong University, Shanghai 200240, China.

[2]Tsung-Dao Lee Institute, Shanghai Jiao Tong University, Shanghai, 200240, China.

[3]Research Center for Electronic and Optical Materials, National Institute for Materials Science, 1-1 Namiki, Tsukuba 305-0044, Japan.

[4]Research Center for Materials Nanoarchitectonics, National Institute for Materials Science, 1-1 Namiki, Tsukuba 305-0044, Japan.

* These authors contributed equally to this work

†Corresponding author. Email: chenguorui@sjtu.edu.cn

**Abstract:** Degeneracies in multilayer graphene, including spin, valley, and layer degrees of freedom, are susceptible to Coulomb interactions and can result into rich broken-symmetry states. In this work, we report a ferromagnetic state in charge neutral ABCA-tetralayer graphene driven by proximity-induced spin-orbit coupling from adjacent $WSe_2$. The ferromagnetic state is further identified as a Chern insulator with Chern number of 4, and its Hall resistance reaches 78% and 100% quantization of $h/4e^2$ at zero and 0.4 tesla, respectively. Three broken-symmetry insulating states, layer-antiferromagnet, Chern insulator and layer-polarized insulator and their transitions can be continuously tuned by the vertical displacement field. Remarkably, the magnetic order of the Chern insulator can be switched by three knobs, including magnetic field, electrical doping, and vertical displacement field.

**Main Text:**

Rhombohedral-stacked multilayer graphene exhibits highly flat conduction and valence bands in the vicinity of the charge neutral point (CNP), where low-energy electrons can be approximately described by a two-band model with an energy-momentum dispersion relation of $E \sim k^N$ (where $E$ represents energy, $k$ represents momentum, and $N$ is the number of layers) (1–5). As a result, graphene multilayers are anticipated to have strong Coulomb interactions (6–9). Additionally, the low-energy bands in graphene are largely associated with momentum-space Berry curvatures and exhibit multiple degeneracies, including spin, valley, and layer degrees of freedom (10, 11). These degeneracies are believed to be susceptible to symmetry-breaking effects induced by interactions (9, 12–15). Consequently, it is predicted that rhombohedral-stacked multilayer graphene can host a diverse range of interaction-driven broken-symmetry states, including the anticipated Chern insulator phase when the top and bottom layers possess opposite valley flavors (7, 9, 14–16). Recent success of fabricating high-quality rhombohedral-stacked multilayer graphene on hexagonal boron nitride (hBN) devices provides promising opportunities to investigate the intriguing broken-symmetry states (17–20).

When the layer number of graphene increases to four, the Coulomb interactions become sufficiently strong to spontaneously break symmetries, leading to layer-resolved charge distribution (layer pseudospin polarization) associated with four spin-valley flavors, in charge neutral ABCA-tetralayer graphene (ABCA-4LG) on hBN. Recent experiment reported a layer-antiferromagnetic (LAF) insulator, in which the flavors of $(K,\uparrow)$ and $(K',\uparrow)$ are polarized at the top layer, while $(K,\downarrow)$ and $(K',\downarrow)$ at the bottom layer ($K$ and $K'$ correspond to two valleys, $\uparrow$ and $\downarrow$ correspond to two spins), in crystalline ABCA-4LG and ABCAB-pentalayer graphene (20–22). Such LAF insulating state is absent in ABC-trilayer on hBN (19, 23). By applying a large vertical displacement field, one can manipulate the charge distribution of the flavors. Consequently, one expects the emergence of partial and full layer-charge polarizations, which are associated with quantum anomalous Hall (QAH) and quantum valley Hall (QVH), respectively. Indeed, a continuous phase transition from LAF under balanced layer-charge polarization to layer polarized insulator (LPI, also referred as QVH according to certain theory and experimental results (14, 15, 24)) under full layer-charge polarization were observed when increasing displacement field (20). However, quantum anomalous Hall, namely Chern insulator, under partial layer-charge polarization, was absent experimentally.

In this report, we present ferromagnetism in charge neutral ABCA-4LG by introducing spin-orbit coupling (SOC) from an adjacent layer of $WSe_2$. We observe an anomalous Hall hysteretic loop, exhibiting a significant Hall resistance $R_{xy} = 5$ k$\Omega$ at zero magnetic field. The $R_{xy}$ value rapidly quantizes to $h/4e^2$ (6.4 k$\Omega$) at a remarkably low magnetic field of 0.4 T, following the Streda formula of a Chern number $C = 4$, by which the ferromagnetic state is evidenced to be a high-order Chern insulator (25). By increasing the vertical displacement field $D$ at CNP, we continuously achieve three distinct broken-symmetry insulating states: LAF near zero $D$, Chern insulator at intermediate $D$, and LPI at large $D$. This observation highlights the electrical tuning of magnetic orders originated from the balanced, partial and full layer pseudospin polarizations. Comparatively, when examining ABCA-4LG without $WSe_2$, we only observe LAF and LPI states. We propose that the nontrivial band topology and long-range magnetism originate from the interplay of intrinsic strong Coulomb interactions in the flat bands of ABCA-4LG and proximity-induced SOC from $WSe_2$. Furthermore, we demonstrate the ability to manipulate the magnetic order through three tuning knobs, magnetic field, doping, and vertical displacement field, revealing the orbital magnetism characteristic of the Chern insulator and its tunable broken-symmetries.

The presence of ABCA-4LG domains in exfoliated tetralayer graphene flakes is confirmed through scanning near-field infrared microscopy (SNOM) (fig. S1). To stabilize its stacking during subsequent fabrication processes, the ABCA-4LG domain is isolated from adjacent ABAB domains by cutting using an atomic force microscope. Subsequently, the ABCA-4LG domain is encapsulated between exfoliated hBN thin films, with a monolayer of $WSe_2$ added between ABCA-4LG and the top hBN layer. The resulting heterostructure is fabricated into a Hall bar geometry, featuring one-dimensional edge contacts, a metal top gate, and a doped silicon bottom gate. Throughout the fabrication procedures, the stacking order of ABCA-4LG under hBN coverage is monitored using the phonon-polariton assisted near-field optical imaging technique (fig. S2D) (20). Figure 1A shows the optical image of the device, and Fig. 1B provides a schematic representation (see materials and methods for further details on WSe$_2$ crystal growth and device fabrication). The top and bottom gates enable individual tuning of the doping $n$ and the vertical displacement field $D$ applied to the ABCA-4LG (materials and methods S3).

Transport measurements are conducted on ABCA-4LG both with and without WSe$_2$. In the case of ABCA-4LG without WSe$_2$, as shown in fig. S3A, two peaks in the longitudinal resistance $R_{xx}$ are observed at CNP for $D = 0$ and large $|D|$. These peaks correspond to the interaction-driven LAF insulator and LPI states, respectively (20, 21). Notably, a low-resistance region near $|D| \sim 0.1$ V/nm connects these two insulators, indicating a gap closure during the continuous phase transition from LAF to LPI. The Hall resistance $R_{xy}$ at a magnetic field of -0.5 T in fig. S3B exhibits the expected sign change across the CNP for the entire range of $D$. In contrast, in ABCA-4LG with WSe$_2$ (Fig. 1C), $R_{xx}$ displays similar features of LAF and LPI at the CNP for both zero and large $D$. However, in the $R_{xy}$ measurement shown in Fig. 1D, at intermediate values of $D$ near $\pm 0.1$ V/nm, the sign change of $R_{xy}$ shifts towards the positive $n$ side, resulting in a prominent $R_{xy}$ at the CNP. Considering that the only difference between these two devices is the presence of WSe$_2$, the distinct nonzero $R_{xy}$ at the CNP is attributed to WSe$_2$.

To investigate the large $R_{xy}$ at CNP for intermediate $D$ in ABCA-4LG with WSe$_2$, we measured its magnetic field dependence. Figure 1E clearly shows the hysteretic anomalous Hall effect, providing clear evidence for the ferromagnetism in ABCA-4LG with WSe$_2$. As a comparison, when we performed the same measurements on ABCA-4LG without WSe$_2$, no ferromagnetic behavior was observed under the same conditions. It is believed that significant SOC on the order of milli electron volts (meV) can be introduced into graphene through proximity with WSe$_2$ due to the hybridization between electron wave functions of both crystals (26, 27). The hysteresis of $R_{xy}$ disappears as the temperature increases to 11 K (Fig. 1E), which is consistent with the calculated and experimental estimates of the proximity-induced SOC strength of about 1 meV in twisted or Bernal bilayer graphene with WSe$_2$ (28–31). While the intrinsic SOC in graphene was predicted to stabilize a two-dimensional topological insulator known as quantum spin Hall by Kane and Mele (32), in practice it is negligibly weak for experimental realization (on the order of micro electron volts) (33–35). However, the proximity-induced SOC in graphene is significant and has been predicted to give rise to intriguing topological phases in multilayer graphene (36, 37).

To explore the topological phase in ABCA-4LG with SOC, we further cool down the sample to $T = 0.1$ K, and perform measurements of $R_{xy}$ and $R_{xx}$ as a function of $n$ and $B$ at $D = -0.1$ V/nm. Inset of Fig. 2C displays the hysteresis of $R_{xy}$, highlighting that the $R_{xy}$ value increases to 5 kΩ at zero magnetic field at $T = 0.1$ K. In Fig. 2B, the large $R_{xy}$ signal persists over a range of approximately -0.1 to 0.4×10$^{12}$ cm$^{-2}$ and exhibits a sharp sign reversal near zero magnetic field due to the anomalous Hall effect. We need to note that the value of $n$ is calculated from gate voltages using a parallel plate capacitor model (materials and methods S3), which may introduce quantitative inaccuracies when graphene becomes insulating at CNP. At remarkably low fields of 0.4 T, the $R_{xy}$ is rapidly quantized at 6.4 kΩ, corresponding to a quantum Hall resistance of $h/4e^2$, following the Streda formula $n = \nu eB/h$ for $\nu = 4$. The presence of $\nu = 4$ quantum Hall state is further evidenced by the corresponding $R_{xx}$ fan diagram shown in Fig. 2A, where a minimum in $R_{xx}$ starts to develop along the slope of the $\nu = 4$ quantum Hall states, represented by the dashed lines.

We argue that the observed quantum Hall state in the presence of a magnetic field is, in fact, a manifestation of the quantum anomalous Hall effect in a Chern insulator with a Chern number $C = 4$. Firstly, when sweeping the magnetic field back and forth within a small range of ±0.2 T at CNP and $D = -0.1$ V/nm, we clearly observe an anomalous Hall signal exhibiting a ferromagnetic hysteresis loop. The measured anomalous Hall signals of $R_{xy} = 5.0$ kΩ and 3.5 kΩ for up and down sweeps of the magnetic field, respectively, correspond to 78% and 55% of the quantized $R_{xy}$ value of $h/4e^2$. We emphasize that the data presented in the main text are displayed in their raw form, without undergoing any data processing such as symmetrization or anti-symmetrization. The emergence of the anomalous Hall signal signifies the breaking of time-reversal symmetry and the presence of ferromagnetism, both of which are crucial hallmarks of the Chern insulator (38–40). Secondly, both the resistance $R_{xx}$ and the conductivity $\sigma_{xx}$ at zero magnetic field trend to approach zero while lowering the temperature (fig. S4), unambiguously demonstrate the quantum Hall type insulating behavior at zero magnetic field. This is indicative of the presence of an exchange gap which serves as another signature of the Chern insulator. At the same time, the temperature dependence of $R_{xy}$ suggests that the incomplete quantization of Hall resistance at zero magnetic field may be attributed to the sample quality. Thirdly, only $v = 4$ quantum Hall state emerges within the range of ±2 T, which is a characteristic feature of a quantized anomalous Hall state. Otherwise, additional quantum Hall states would be expected to develop. Indeed, when tuning $D$ outside the range of the anomalous Hall region, a series of quantum Hall states emerge as shown in fig. S5. Lastly, the Hall angle at zero magnetic field, defined as $R_{xy}/R_{xx}$, exhibits a remarkably large value of 3.3, indicating an intrinsic mechanism for the anomalous Hall and, consequently, the nontrivial band topology.

Following the discussions on the Chern insulator, we present the magnetic field-stabilized quantum anomalous Hall (QAH) effect in Fig. 2C and D. The longitudinal and Hall resistances are plotted as a function of the magnetic field along the dashed lines in Fig. 2A and B, respectively, in accordance with the Streda formula for a Chern number with $C = ±4$. The sign of the Chern number is reflected in the sign of $R_{xy}$ and can be switched by controlling the magnetic field. The ferromagnetic hysteric behavior for other values of $n$ and $D$ are shown in fig. S6.

Next, by examining the broken-symmetry states and their transitions along tuning the displacement field for charge-neutral ABCA-4LG with and without WSe$_2$, we can gain valuable insights into the emergence of the Chern insulator. In the case of ABCA-4LG without WSe$_2$, as shown in Fig. 3A (and fig. S3), two broken-symmetry correlated insulating states, LAF and LPI, are observed at CNP. LAF emerges when $D = 0$, where two spin-valley flavors, $(K,↑)$ and $(K',↑)$, are polarized in the top layer, while the other two flavors, $(K,↓)$ and $(K',↓)$, are polarized in the bottom layer. At large values of $D$, all four spin-valley flavors become polarized in the same layer, resulting in the LPI state. Each spin-valley flavor pair corresponds a band with a Chern number of $±N/2$, where $N$ (equals to 4 for tetralayer) represents the number of graphene layers. The sign of the Chern number depends on both the valley label and the mass term (the sense of layer polarization) in the two-band model of ABCA-4LG (*10, 14*). For simplify, the sign can be conveniently represented by the valley-layer locking: a positive sign when $K$ ($K'$) is polarized in the top (bottom) layer and a negative sign when $K$ ($K'$) is polarized in the bottom (top) layer. Following this criterion, for LAF, depicted as the gray diagram in Fig. 3C, the Chern number of $(K,↑)$, $(K',↑)$, $(K,↓)$ and $(K',↓)$ are 2, -2, -2 and 2, respectively, resulting in a zero total Chern number and indicating a topological trivial state. Similarly, for LPI, illustrated as the blue diagram in Fig. 3C, $(K,↓)$ and $(K',↓)$ are transferred from the bottom layer to the top layer, with the signs of their Chern numbers reversing and becoming 2 and -2, respectively. As a result, the total Chern number remains zero for LPI. In the intermediate region between LAF and LPI, $R_{xx}$ continuously decreases to as low as ~2 kΩ and exhibits metallic temperature dependence at low temperatures (fig. S3C). Additionally, no anomalous Hall effect is observed, indicating a gap closure at $D \sim 0.1$ V/nm. These broken-symmetry insulating states in ABCA-4LG without WSe$_2$ are driven by strong Coulomb interactions in the intrinsic flat bands of ABCA-4LG, which allows the breaking of time-reversal symmetry ($\mathcal{T}$) and inversion symmetry ($I$), while preserving the valley-Ising symmetry ($\mathcal{Z}_2$) (*14, 41*), leading to the absence of Chern insulator.

In comparison, Fig. 3B shows $R_{xy}$ as a function of $D$ at CNP for ABCA-4LG with WSe$_2$. Despite that the expected $R_{xy}$ value at CNP should be infinite and experimentally inaccessible, the measured $R_{xy}$ is large for the insulating states, possibly due to the pick-up signal from the large $R_{xx}$. As $D$ is swept, the LAF and LPI states are not significantly affected by SOC due to their large energy gaps (*20, 21*). However, SOC plays a crucial role in the intermediate state between LAF and LPI, leading to the emergence of a Chern insulator state. The Chern insulator can be attributed to a band inversion of one flavor pair, driven by the proximity-induced SOC. At intermediate $D = 0.1$ V/nm, shown as orange diagram in Fig. 3C, $(K',↓)$ is transferred from the bottom layer to the top layer, resulting in a sign change of its Chern number from -2 to 2. This sign change contributes to a total Chern number of 4 when accumulating the Chern numbers of all four flavor pairs. Besides breaking the time-reversal symmetry and inversion symmetry, the interplay of SOC and strong Coulomb interactions can further break the $\mathcal{Z}_2$ symmetry, driving the formation of the Chern insulator state.

The observed Chern number $C = 4$ in ABCA-4LG with WSe$_2$ is compatible with the predicted flavor ferrimagnetic (Fi) state for partial layer-charge polarization in rhombohedral multilayer graphene. Such state have associated "ALL" quantum Hall phases including quantum anomalous charge, spin, valley, and spin-valley Hall state (fig. S8) (*14*, *42*). Our transport measurements primarily detect the quantum anomalous charge Hall effect in ABCA-4LG with SOC, indicating a promising system for further exploration of quantum anomalous spin, valley, and spin-valley Hall effects using spin/valley-sensitive probes. It is worth noting that there are four possible layer-flavor polarization configurations for the "ALL" state at a given $D$ (as shown in fig. S8 for negative $D$ side), we got the configuration in Fig. 3 from the sign of the Chern number and the relation between $n$ and $B$ from the Streda formula.

Additionally, the magnetic order, represented by the sign of the anomalous Hall resistance ($\Delta R_{xy}^B = R_{xy}^{B\uparrow} - R_{xy}^{B\downarrow}$), exhibits intriguing dependence on $n$. In Fig. 4A, a sign reversal of $\Delta R_{xy}$ is observed when $n$ slightly crosses CNP at around $0.28 \times 10^{12}$ cm$^{-2}$, indicating a switching of the magnetic order as the Fermi level crosses the band gap. Figure 4B presents representative line cuts from Fig. 4A at different $n$ values, illustrating the sign reversal of $\Delta R_{xy}^B$ and switching of magnetic order.

Remarkably, the magnetic order can be directly switched by tuning $n$ in fixed $B$. At fixed magnetic fields ranging from 0.28 T to -0.26 T, hysteresis loops of $R_{xy}$ in the $n$-axis are observed, indicating the switching of the magnetic order as $n$ is swept back and forth (Fig. 4D). The sign reversal of $\Delta R_{xy}^n = R_{xy}^{n\uparrow} - R_{xy}^{n\downarrow}$ occurs when the magnetic field crosses zero (Fig. 4C). This demonstrates that the magnetic order of the Chern insulator at CNP can be individually controlled by $B$ and $n$, as sketched in Fig. 4E. The electrical tunability of the magnetic order can be described by an orbital Chern insulator, where the net magnetization arises from two parts, namely the bulk and the protected edge-states. The net magnetization reverses as a result of a significant contribution of the edge-state with opposite signs when doping with different types of charge carriers (*43*). It is worth noting that similar electrical switching of the magnetic order in Chern insulators has been reported in twisted graphene systems, where the large area of the moiré unit cell accounts for the substantial edge-state contribution compared to the bulk contribution (*44–46*). However, in ABCA-4LG, such an argument is not applicable considering the much smaller unit cell of crystalline ABCA-4LG, suggesting unique characteristics of the observed electrical switching of the magnetic order.

Furthermore, the magnetic order also exhibits an interesting dependence on the displacement field. Fig. 3B (lower panel) shows hysteresis loops of $R_{xy}$ in $D$ at the Chern insulator states for $D = \pm 0.1$ V/nm. When sweeping $D$ from negative to positive, $R_{xy}$ is negative near $D = -0.1$ V/nm and positive near $D = 0.1$ V/nm. Conversely, when sweeping $D$ from positive to negative, $R_{xy}$ becomes negative near $D = 0.1$ V/nm and positive near $D = -0.1$ V/nm. However, at a magnetic field of $B = 10$ mT, unlike the sign reversal observed during doping switching, $R_{xy}$ remains positive for both sweeping directions on both sides of $D$ (fig.S7B). These distinct dependences of $R_{xy}$ on $n$ and $D$ can be represented in the parameter space of ALL state in fig. S8.

At last, we note that from the predicted ALL state, the ferromagnetism and therefore the Chern number can be switched by three knobs, $B$, $n$ and $D$, represented by a three-dimensional parameter space shown in fig. S8. We indeed observed the magnetic order switching by these three knobs, which can be demonstrated by corresponding line cuts in such a parameter space. The key to achieve Chern insulator in ABCA-4LG is proximity-induced SOC and strong Coulomb interactions, which should also be expected in thicker rhombohedral multilayer graphene with WSe$_2$, and therefore multilayer graphene with WSe$_2$ can provide a platform of layer number dependent Chern insulator.

Our results showcase the SOC-driven topological orders by van der Waals proximity, and potential of crystalline rhombohedral multilayer graphene as a versatile and easily adjustable platform for investigating the interplay of different interactions. They also open up avenues for investigating topological phase transitions and exploring novel topological phases, such as multiferroic Chern insulators and fractional Chern insulators. The layer-dependent Chern numbers in rhombohedral multilayer graphene provide a valuable natural-crystal resource of high-order Chern insulators for multi-channel quantum computing.

**Acknowledgments:** We thank Fengcheng Wu, Zhenhua Qiao, and Jeil Jung for helpful discussions.

**Funding:** This work is supported from National Key Research Program of China (grant nos. 2021YFA1400100, 2020YFA0309000, 2021YFA1202902), NSF of China (grant nos.12174248, 12074244), and SJTU NO. 21X010200846. G.C. acknowledge the sponsorship from Yangyang Development Fund. K.W. and T.T. acknowledge support from the JSPS KAKENHI (Grant Numbers 20H00354, 21H05233 and 23H02052) and World Premier International Research Center Initiative (WPI), MEXT, Japan.

**Author contributions:** G.C. conceived and supervised the project. Y.S. and K.L. fabricated the devices with and without $WSe_2$, respectively. Y.S., J.Z. and K.L. performed the transport measurements. J.Z. set up the dilution refrigerator. H.D. and R.Z. grew $WSe_2$ bulk crystals. K.W. and T.T. grew hBN bulk crystals. Y.S., J.Z., K.L. and G.C. analyzed the data. Y.S. and G.C. wrote the paper with input from all authors.

**Competing interests:** Authors declare that they have no competing interests.

**Data and materials availability:** All data in the main text or the supplementary materials are available from the corresponding authors upon reasonable request.

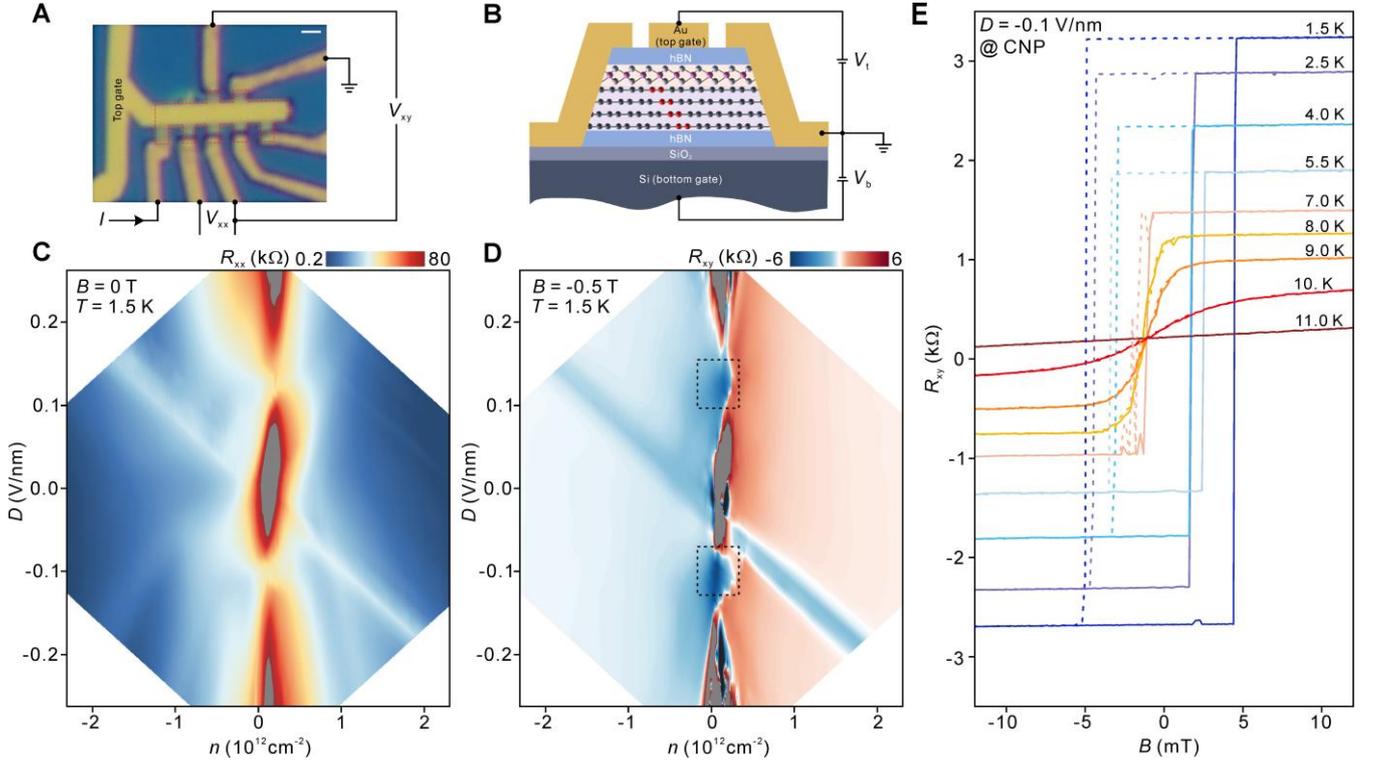

**Fig.1. Schematic and transport of ABCA-4LG with WSe$_2$.** (**A**) Optical image of the hBN/WSe$_2$/ABCA-4LG/hBN device, including a schematic of the transport measurement configuration. The Hall bar shaped graphene is highlighted by red dash line. Scale bar, 1 μm. (**B**) Schematic side view of a dual gate WSe$_2$/ABCA-4LG device. The crystal structures of ABCA-4LG and WSe$_2$ monolayer are shaded by light purple and light orange, respectively. A unit cell of ABCA-4LG is labeled in red. (**C, D**) Color plot of longitudinal $R_{xx}$ (C) and Hall $R_{xy}$ (D) resistance at $T$ = 1.5 K as a function of carrier density $n$ and displacement field $D$. $R_{xx}$ is measured under zero magnetic field, showing insulating states at $D$ = 0 V/nm and large |$D$|. $R_{xy}$ is measured at a low magnetic field, $B$ = -0.5 T. The dashed rectangles near $D$ = ±0.1 V/nm in (D) outline the region of Chern insulator, where the sign change of $R_{xy}$ shifts towards the positive $n$ side, resulting in a significant $R_{xy}$ at the CNP. The gray and black regions represent larger and smaller values than the color scales, respectively. (**E**) Hysteretic loops of anomalous Hall signals at CNP and $D$ = -0.1 V/nm at various temperatures above $T$ = 1.5 K.

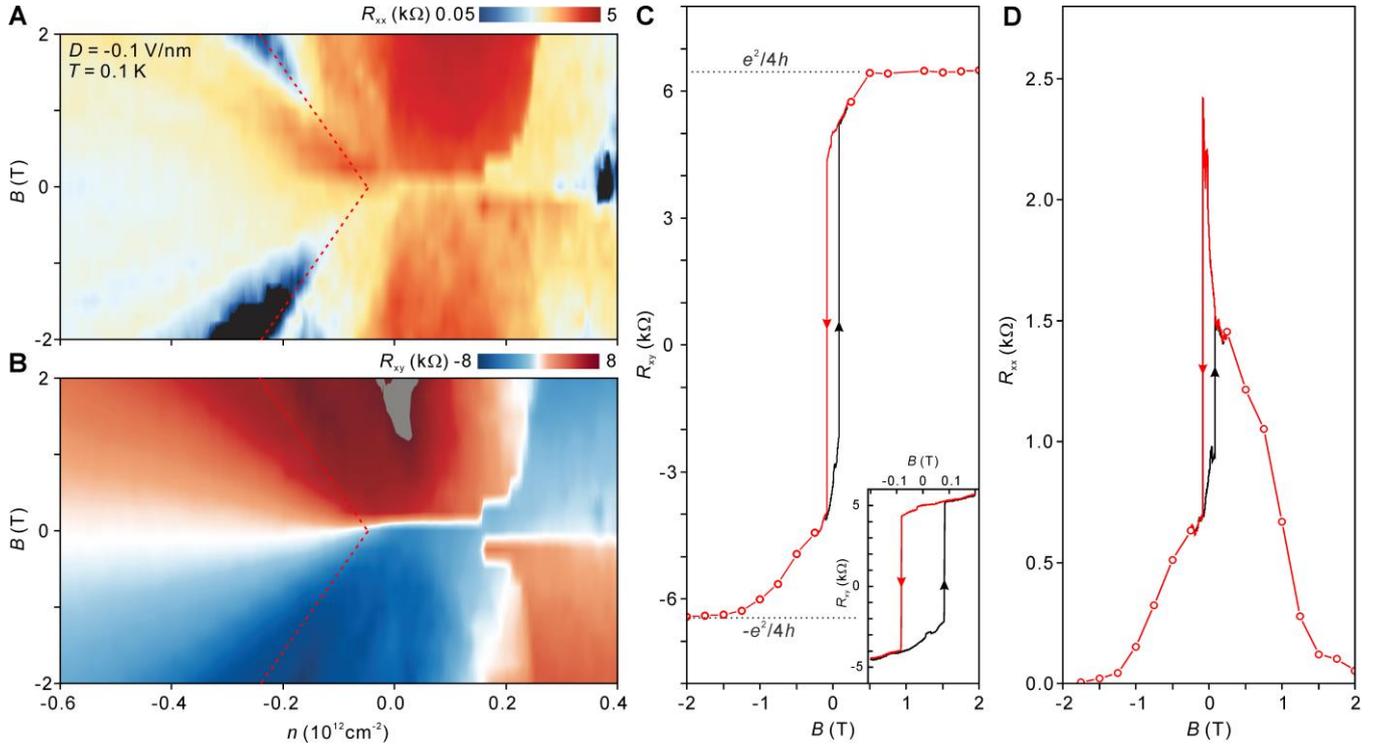

**Fig.2. Chern insulator and magnetic-field-stabilized QAH effect at $D = -0.1$ V/nm. (A, B)** Color plot of $R_{xx}$ (A) and $R_{xy}$ (B) as a function of carrier density $n$ and magnetic field $B$ at $D = -0.1$ V/nm and $T = 0.1$ K. The $\nu = 4$ quantum Hall state can be identified following the Streda formula (denoted by the red dashed lines in (A) and (B)). Near zero magnetic field, $R_{xy}$ exhibits a sharp sign reversal, indicating the presence of the anomalous Hall effect. Field is swept from positive to negative (namely, sweeping down) in both plots. **(C, D)** Magnetic-field-dependent $R_{xx}$ and $R_{xy}$ at $D = -0.1$ V/nm. Data within a small range of $B = \pm 0.2$ T are collected from continuous $B$ sweeping up and down (see inset of (C) for zooming in). At higher $B$ fields, data points (empty circles) of $R_{xx}$ and $R_{xy}$ are acquired along the dashed lines in (A) and (B) following the Streda formula.

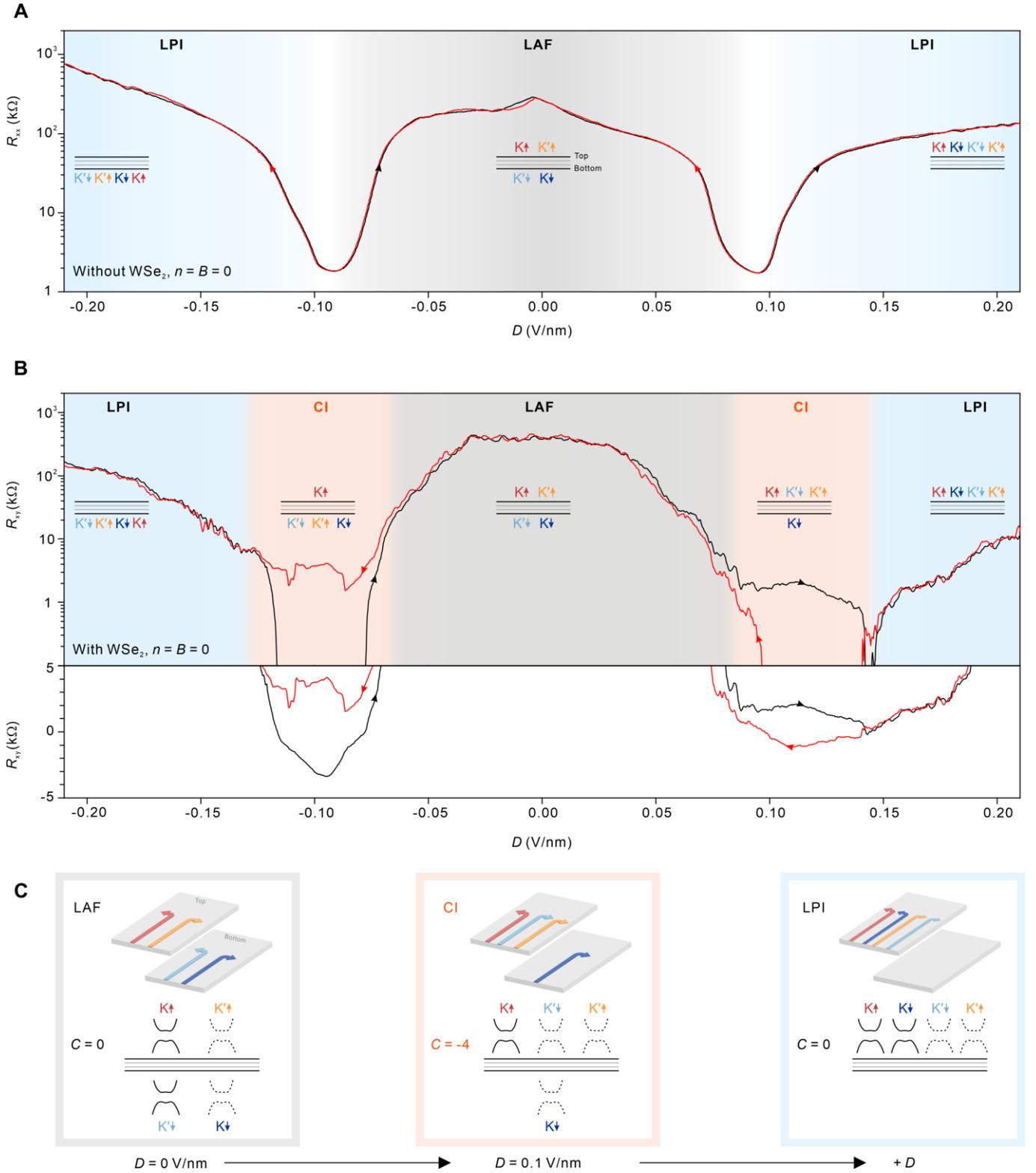

**Fig.3. Broken-symmetry states at CNP in ABCA-4LG with and without SOC.** (**A**) $R_{xx}$ as a function of $D$ at $B = 0$ T in ABCA-4LG without SOC (WSe$_2$). Gray and blue shaded regions correspond to the interaction-driven layer antiferromagnetic (LAF) insulator and layer polarized insulator (LPI) states, respectively. (**B**) $R_{xy}$ as a function of $D$ at $B = 0$ T in ABCA-4LG with SOC (WSe$_2$). In addition to LAF and LPI states, the Chern insulator (CI) states are observed with hysteresis loops at $D = \pm 0.1$ V/nm (orange shaded regions in the upper panel). The lower panel shows a zoomed-in plot of $R_{xy}$ in linear scale. Inset schematics in (A) and (B) indicate the layer polarization of spin-valley flavors for different broken-symmetry states (for the CI state, only one representative polarization phase is presented. The completed phase diagram is shown in Fig. S8). The black and gray lines in the inset represent four graphene layers. Valleys and spins are represented by $K$, $K'$, and ↓, ↑. (**C**) Schematics of the Chern numbers and Hall conductivity

contributions of four spin-valley flavor pairs for different broken-symmetry states. Each spin-valley flavor pair corresponds to a band with a Chern number of ±2, denoted by solid (+2) or dashed (-2) bands in the lower panels, respectively. The directions (left and right) of the arrows in the upper panels represent the corresponding Hall conductivity contributions ($+2e^2/h$ and $-2e^2/h$).

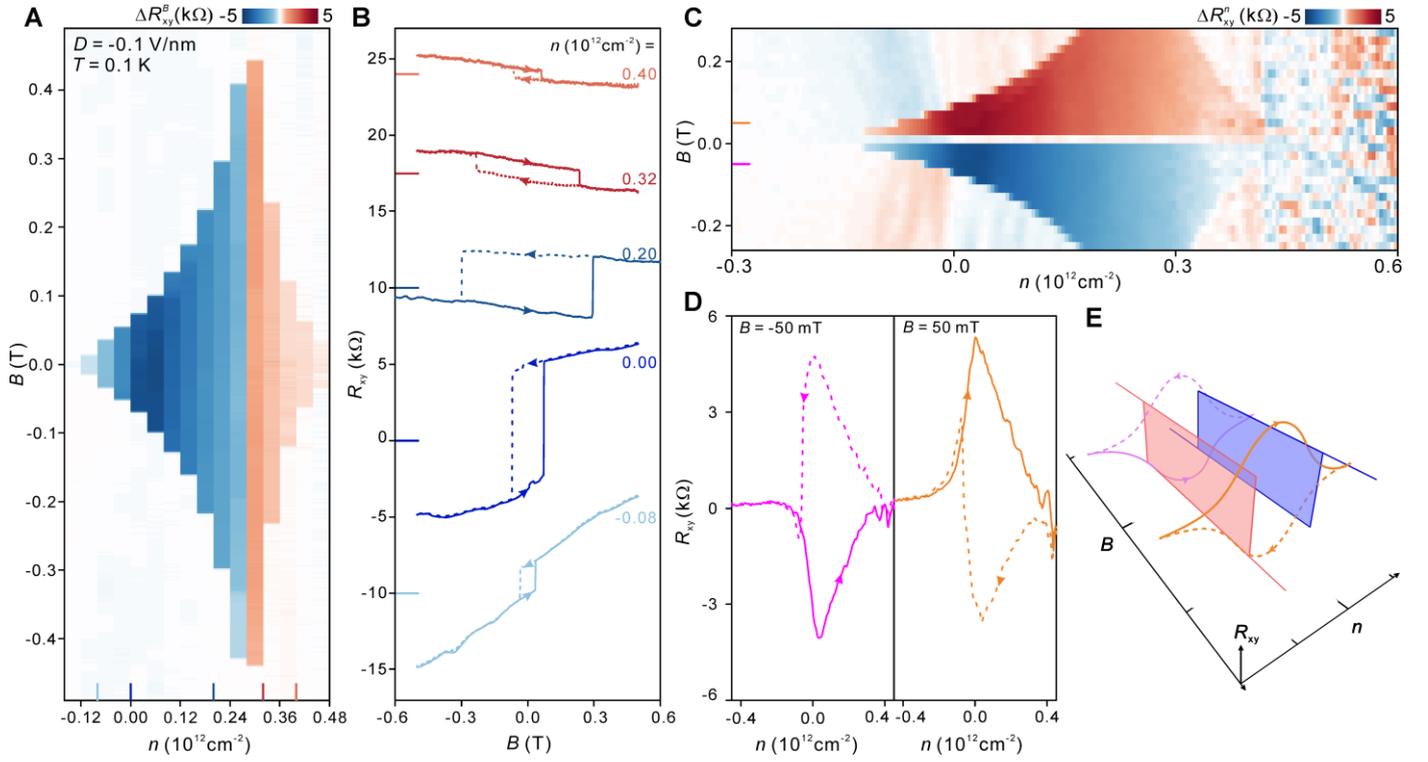

**Fig.4. Electrical switching of the magnetic order.** (**A**) The anomalous Hall resistance when sweeping $B$, defined by $\Delta R_{xy}^{B} = R_{xy}^{B\uparrow} - R_{xy}^{B\downarrow}$ and represented by colors, at different doping near CNP at $D = -0.1$ V/nm and $T = 0.1$ K. An abrupt sign change of $\Delta R_{xy}^{B}$ occurs at around $n = 0.28 \times 10^{12}$ cm$^{-2}$. (**B**) Magnetic hysteresis loops of $R_{xy}$ at $n = 0.4 \times 10^{12}, 0.32 \times 10^{12}, 0.2 \times 10^{12}, 0, -0.08 \times 10^{12}$ cm$^{-2}$, with colors corresponding to the doping positions labeled in (A). Data are vertically offset for clarity. Solid lines on the $y$-axis mark the positions of zero $R_{xy}$ for each curve. (**C**) The anomalous Hall resistance when sweeping $n$, defined by $\Delta R_{xy}^{n} = R_{xy}^{n\uparrow} - R_{xy}^{n\downarrow}$, at fixed $B$ ranging from -0.25 T to 0.25 T. Temperature and $D$ are same as in (A). (**D**) Electrical hysteresis loops of $R_{xy}$ at $B = \pm 50$ mT, with colors corresponding to the magnetic fields labeled in (C). (**E**) Schematic of two individual knobs, $B$ and $n$, of tuning the magnetic order of the Chern insulator at CNP in ABCA-4LG.

## S1. Growth of WSe$_2$ bulk crystal

Single crystals of WSe$_2$ were grown using chemical vapor transport (CVT) method with I$_2$ as the transport agent. Stoichiometric mixtures of niobium slug (PrMat, 99.95%) and selenium granule (PrMat, 99.9999%) with a total mass of about 2 g were loaded to a quartz tube and sealed under vacuum of ~ $5\times10^{-2}$ torr. The quartz tube was then slowly heated to 750 °C overnight to avoid over-pressure. Afterwards, the products and 150 mg of I$_2$ were air-freely sealed in a quartz tube of 18 mm in inner diameter and 20 cm in length. A two-zone furnace was used with a temperature gradient of 980 °C $\rightarrow$ 880 °C. The furnace was held at the temperature for 5 days before being cooled to room temperature over 6 hours.

## S2. Device fabrications

Graphene, WSe$_2$, and hBN flakes are mechanically exfoliated from bulk crystals onto SiO$_2$(285 nm)/Si substrates, with their layer numbers determined by optical reflectance contrast. The stacking orders of tetralayer graphene are identified using a scanning near-field optical microscope (SNOM) (Fig. S1). A dry transfer method is employed to stack the desired hBN/WSe$_2$/4LG/hBN heterostructures, followed by a second SNOM characterization to verify the stacking orders of rhombohedral graphene. Subsequently, standard e-beam lithography, reactive ion etching, and metal evaporation are conducted to transform the devices into Hall bar geometry with the one-dimensional edge contacts (*1*) as shown in Fig. S2.

## S3. Transport measurement

Transport measurements above 1.5 K were performed in an Oxford variable temperature insert (VTI) system. Measurements at 0.1 K were performed in a Leiden dilution refrigerator. Stanford Research Systems SR860, SR830 and Guangzhou Sine Scientific Instrument OE1201 lock-in amplifiers were used to measure the sample resistances with a small ac bias current of 1~10 nA at a low frequency (~17.7 Hz). Homebuilt low-pass resistor-capacitor (RC) filters were equipped in the dilution refrigerator to reduce high frequency noise. Gate voltages were applied using Keithley 2400 Source Meters. In our dual gate device, the top and bottom gate voltages $(V_t, V_b)$ independently tune the carrier density $n$ and the vertical displacement field $D$ in tetralayer graphene: $n = (D_b + D_t)/e$ and $D = (D_b - D_t)/2$, where $D_b = \varepsilon_b(V_b - V_b^0)/d_b$, $D_t = \varepsilon_t(V_t - V_t^0)/d_t$. Here, $\varepsilon$ and $d$ represent the dielectric constant and thickness of the dielectric layers, respectively, $V_b^0$ and $V_t^0$ are effective offset voltages caused by intrinsic carrier doping and $e$ is the electron charge.

## S4. Symmetrization and anti-symmetrization of the magneto-transport data

Most of the data presented in this report are in their raw form without any processing, with the exception of the temperature dependance of hysteretic behaviors shown in Fig.S4. We note that finite couplings between longitudinal and transverse directions would mix $R_{xx}$ and $R_{xy}$ signals. According to that $R_{xy}$ is anti-symmetric with respect to magnetic field $B$, whereas $R_{xx}$ is symmetric, we can separate the two components using standard symmetrization and anti-symmetrization procedures to obtain the data shown Fig.S4. The symmetrization procedure for $R_{xx}$ is as follows:

$$\rho_{xx}(B, up) = \frac{R_{xx}(B, up) + R_{xx}(-B, down)}{2} \cdot \frac{W}{L}$$

$$\rho_{xx}(B, down) = \frac{R_{xx}(B, down) + R_{xx}(-B, up)}{2} \cdot \frac{W}{L}$$

The anti-symmetrization procedure for $R_{xy}$ is as follows:

$$\rho_{xy}(B, up) = \frac{R_{xy}(B, up) - R_{xy}(-B, down)}{2}$$

$$\rho_{xy}(B, down) = \frac{R_{xy}(B, down) - R_{xy}(-B, up)}{2}$$

where up/down indicates the direction of the magnetic field sweeping, the channel width $W = 1.1$ um and channel length $L = 1.5$ um. The conductance is defined as: $\sigma_{xx} = \rho_{xx}/(\rho_{xx}^2 + \rho_{xy}^2)$. In Fig.S4, we use $\rho_{xx}$, $\rho_{xy}$, and $\sigma_{xx}$ to represent the symmetrized or anti-symmetrized data.

**Fig. S1.**

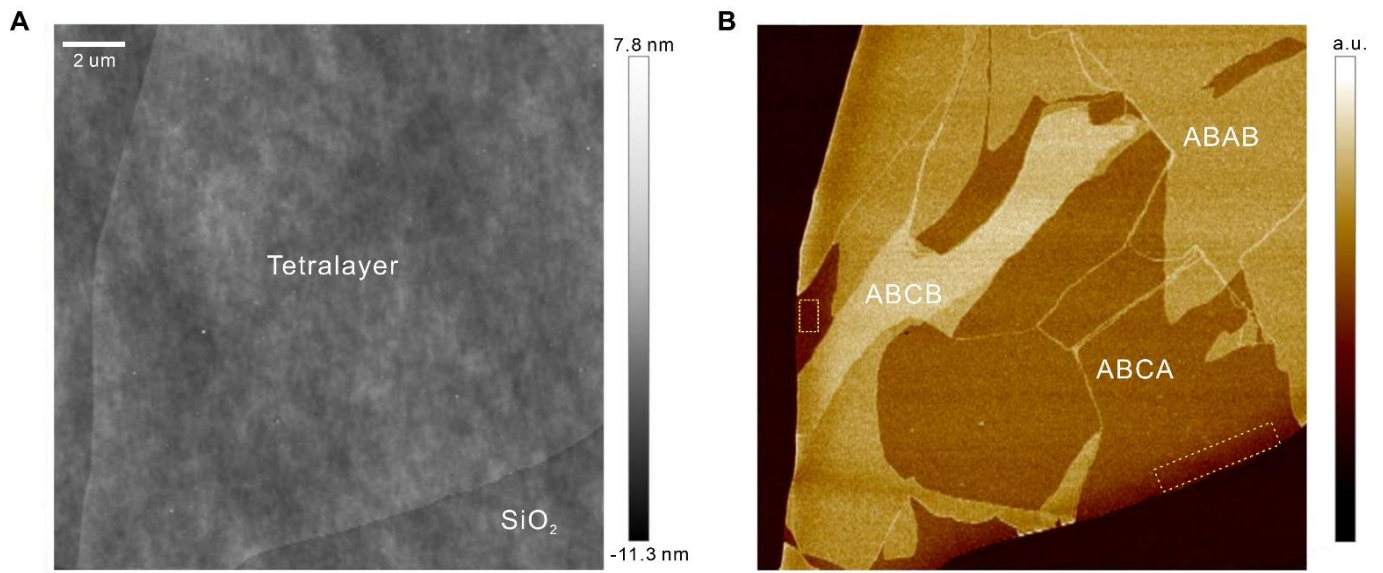

Fig.S1. **Identification of ABCA-4LG.** **(A)** Atomic force microscope topography image of an exfoliated tetralayer graphene on SiO$_2$/Si. **(B)** Near-field infrared image corresponding to (A). Different domains, ABAB, ABCB and ABCA can be clearly distinguished according to different contrasts. Under this incident light wavenumber (1587.5 cm$^{-1}$), ABCB is the brightest, and ABCA is the darkest (*2*, *3*). While it should be noted that at the boundary of the sample, contrast tends to darker due to tip disturbance from the step between graphene and the substrate, as showing in yellow dashed box.

**Fig. S2.**

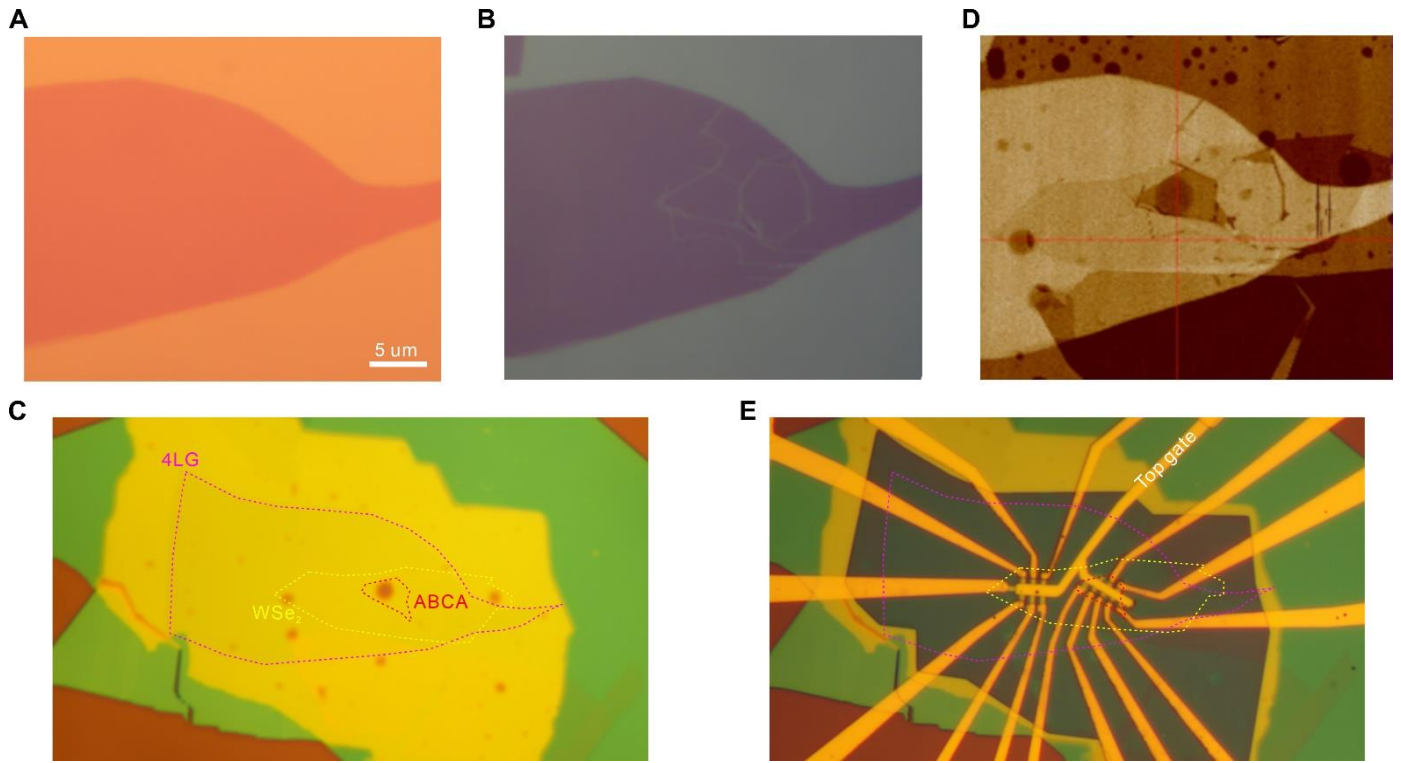

Fig.S2. **Device of ABCA-4LG with WSe₂ on top.** (**A**) Optical image of the tetralayer graphene flake shown in Fig.S1. (**B**) The same image as in (A) but ABCA domains are isolated after anodic-oxidation-assisted AFM cutting (*2*, *3*). (**C**) Optical image of an hBN/WSe₂/graphene/hBN heterostructure prepared using a dry transfer technique. (**D**) SNOM image of the sample, which helps to reconfirming the existence of rhombohedral domains. (**E**) The final ABCA-4LG/WSe₂ device.

**Fig. S3.**

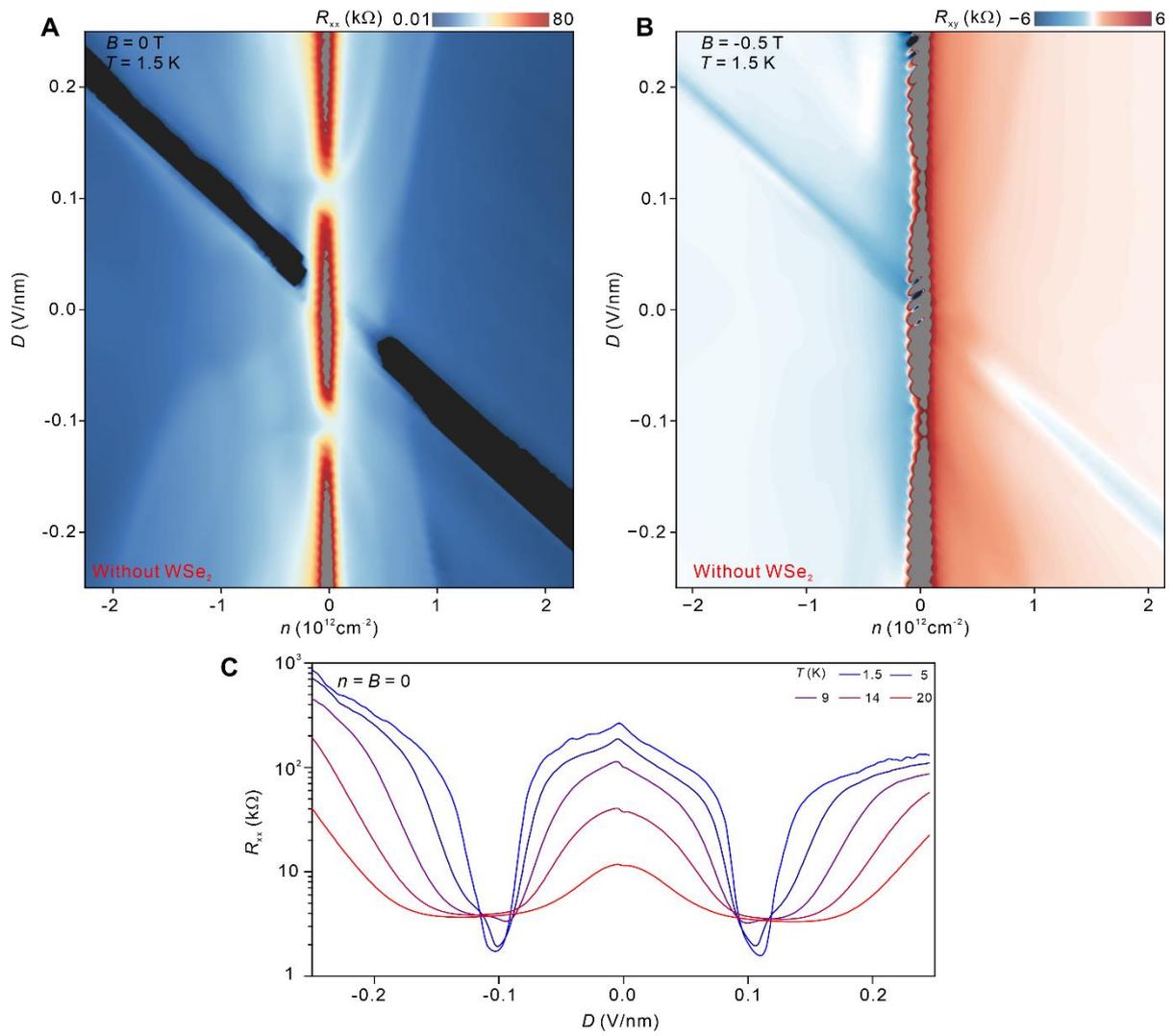

Fig.S3. **Transport on ABCA-4LG without WSe$_2$.** (**A, B**) $R_{xx}$ (A) and $R_{xy}$ (B) as a function of carrier density $n$ and displacement field $D$. The temperature and magnetic field are the same as that in Fig.1(C, D) in the main text. Insulating regions at CNP for $D = 0$ and large $|D|$ are correspond to the LAF and LPI states, respectively. A low-resistance region near $|D| \sim 0.1$ V/nm connects these two insulators, indicating a gap closure during the continuous phase transition from LAF to LPI. (**C**) $R_{xx}$ as a function of $D$ for $n = B = 0$ at various temperatures from 1.5 K to 20 K, exhibiting metallic behavior near $|D| = 0.1$ V/nm (*2*).

**Fig. S4.**

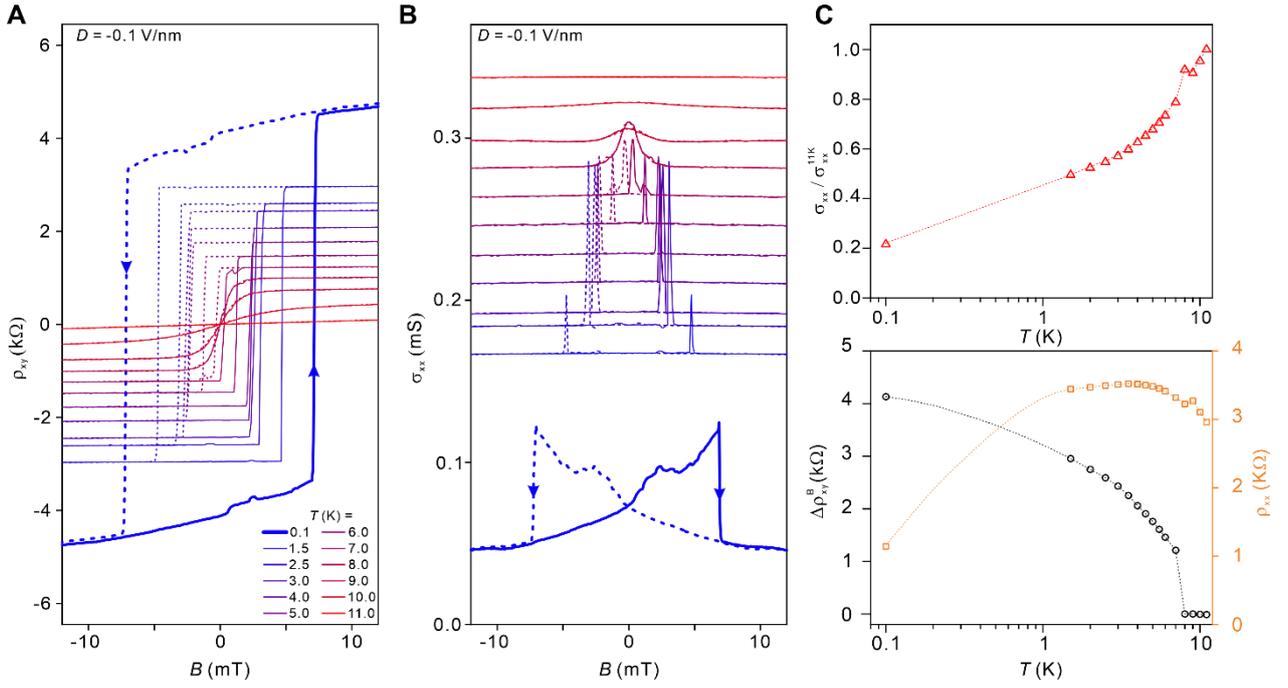

Fig.S4. **Temperature dependence of the hysteresis at $D = -0.1$ V/nm. (A, B)** Magnetic field dependent anti-symmetrized $\rho_{xy}$ and symmetrized $\sigma_{xx}$ at varying temperatures for the ferromagnetic sate at CNP. The coercive field at 0.1 K has been manually reduced by a factor of 100 for clarity. **(C, D)** The evolution of corresponding $\Delta\rho_{xy}^B$, $\rho_{xx}$ and $\sigma_{xx}$ as a function of temperature, with dashed lines are guide to eye. The amplitude of $\Delta\rho_{xy}^B$ decreases with increasing $T$ and vanishes at 8 K, while both $\rho_{xx}$ and $\sigma_{xx}$ decrease with decreasing $T$ at low temperature regime, supporting the identification of the state as a Chern insulator. The data processing of the anomalous Hall signals in this figure is describe in Materials and Methods S4.

**Figure S5.**

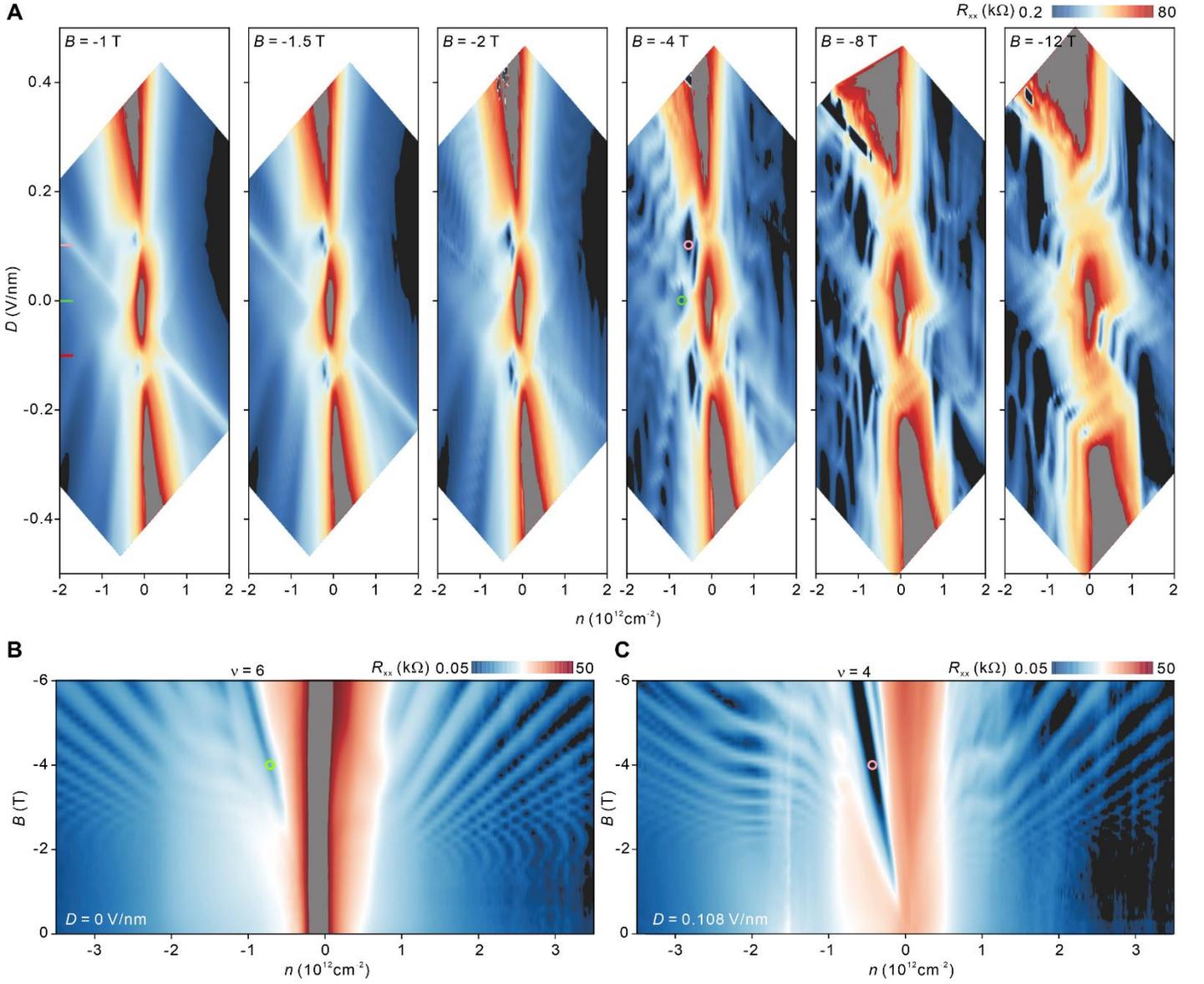

Fig.S5. **Development of quantum Hall states in magnetic fields.** (**A**) $n$-$D$ color plot of $R_{xx}$ at different $B$, measured at $T = 1.5$ K. As increasing the field, $\nu = 4$ state near $|D| = 0.1$ V/nm (pink and red lines) is firstly developed and expands more widely than other quantum Hall states. It is also worth noting that $\nu = 3$ state competes with $\nu = 4$ Chern insulator state in the vicinity of $|D| = 0.1$ V/nm, but it is too weak to be repelled ultimately. At higher $B$, other quantum hall states start to emerge outside the region of the Chern insulator state. (**B, C**) Landau level fan diagram of ABCA-4LG with WSe$_2$ device, measured at $T = 1.5$ K. Data in panel (B) are measured at $D = 0$ V/nm (green line in (A)), where the first developed landau level is identified to be $\nu = 6$. In contrast, data in panel (C) are measured at near $D = 0.1$ V/nm (pink line in (A)). Similar to the case at $D = -0.1$ V/nm (as in main Fig.2A), a robust and prominent $\nu = 4$ state can be observed. Green and pink circles indicate the corresponding locations of these two distinct states in (A) at $B = 4$T.

**Figure S6.**

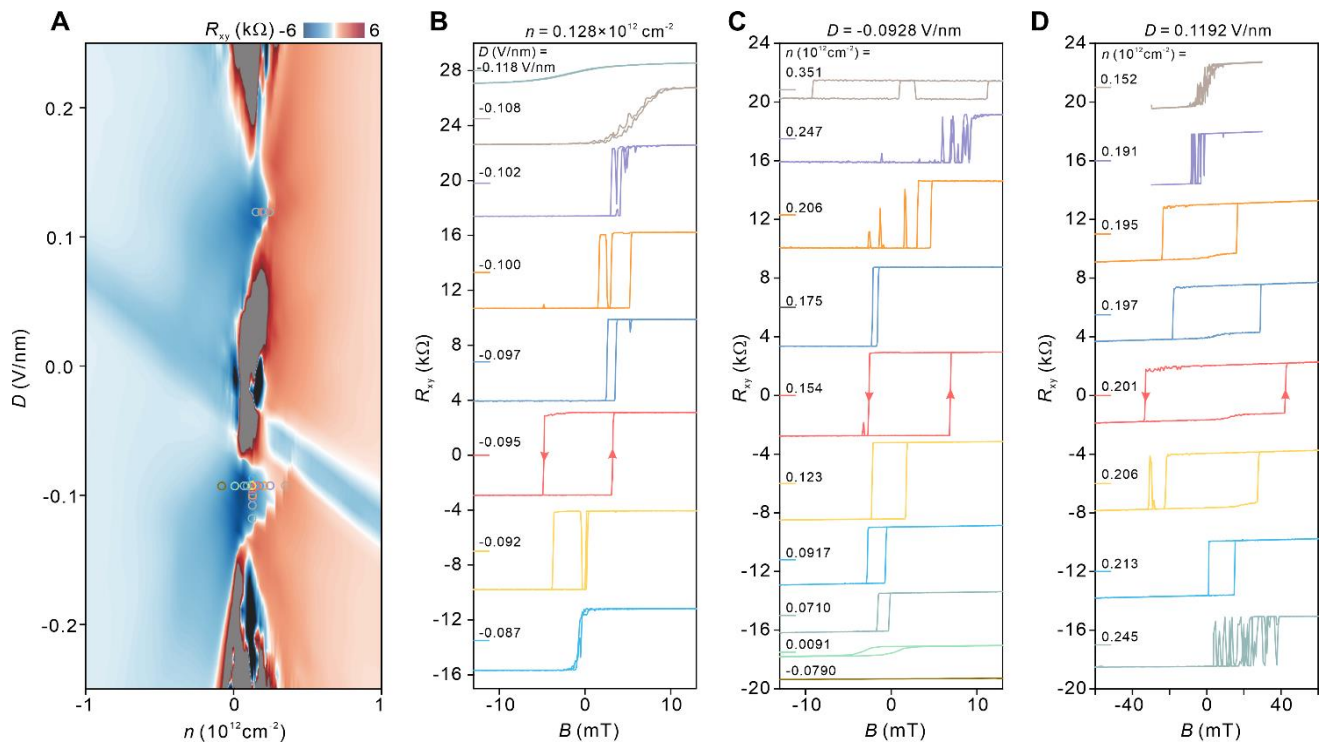

Fig.S6. **Development of anomalous Hall resistance at different *n* and *D* in ABCA-4LG with WSe₂. (A)** *n-D* color plot of $R_{xy}$ at $B = -0.5$ T. The Data are converted from Fig.1D in the main text. **(B, C, D)** $R_{xy}$ loops measured at various *n* and *D*, indicated by the circles with corresponding colors in (A). Panel (B) and (C) display the hysteresis loops at $D < 0$ side. Panel (D) displays the hysteresis loops at $D > 0$ side, with a smaller anomalous Hall signal of approximately 2 kΩ. Therefore, we primarily focus on the $D = -0.1$ V/nm side in this report.

**Figure S7.**

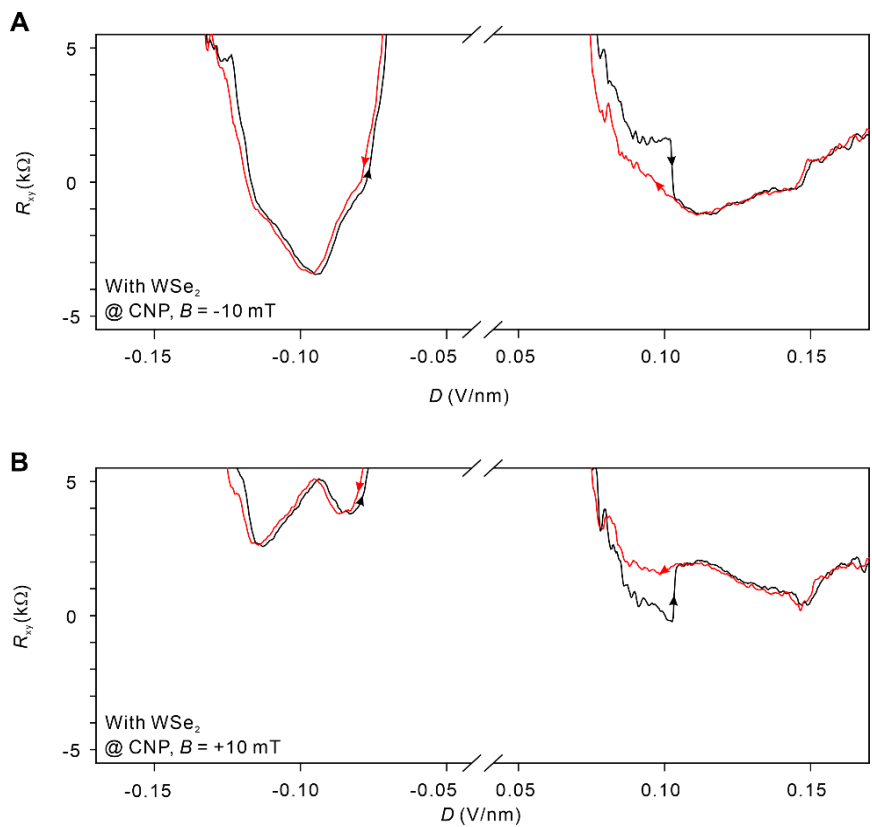

Fig.S7. **Polarized magnetic order on *D* under a small magnetic field. (A, B)** $R_{xy}$ as a function of *D* in a magnetic field. At *B* = -10 mT (A), $R_{xy}$ remains negative for both sweeping directions on both sides of *D*. Conversely, at *B* = 10mT (B), $R_{xy}$ becomes positive for both sweeping directions on both sides of *D*. Compared to the relatively robust dependence of $R_{xy}$ on *n* in finite *B* (Fig.3C), the distinct behaviors observed for the vertical displacement field suggest different mechanisms underlying the switching of the magnetic order by *n* and *D*.

**Figure S8.**

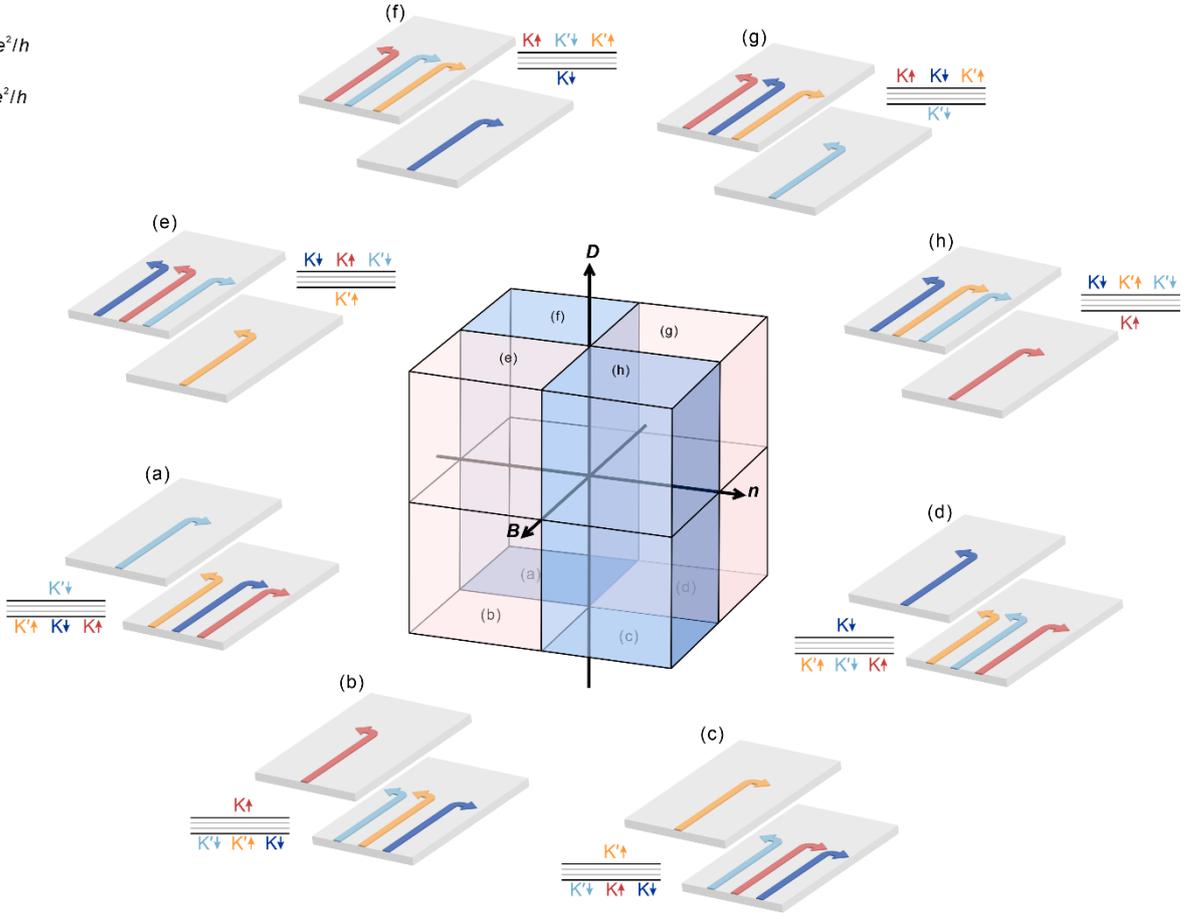

Fig.S8. **Parameter space of "ALL" quantum anomalous Hall phases.** Middle: Schematic of the eight different "ALL" phases in a three-dimensional coordinate, where the x, y, z axis are perpendicular magnetic field $B$, carrier doping $n$ and vertical displacement field $D$, respectively. The colors of light pink and blue represent positive and negative Chern numbers, respectively, corresponding to different signs of Hall conductance $\sigma_{xy}$ (top left). This is a manifestation of the switchable orbital magnetization. The lowercase letters from (a) to (h) in each quadrant label the "All" phases depicted in the surrounding schematics. For instance, at $n, D < 0, B > 0$, the spin and valley flavors of the Chern insulator phase spontaneously polarize in the way described in (b): $(K',\downarrow)$ $(K',\uparrow)$ and $(K,\downarrow)$ reside in the bottom layer, while $(K,\uparrow)$ is in the top layer. Note that the back scatterings caused by edge roughness can gap a pair of counter-propagating edge states in the bottom layer and each of the left two edge states contributes $Ne^2/2h$ to the Hall conductivity (4, 5), that is $2e^2/h$ for tetralayer graphene ($N = 4$). Thus, the net Hall conductance is expected to be $\sigma_{xy} = +4e^2/h$ with Chern number $C = 4$, which agrees with our experiment (Fig.2). The schematics also demonstrates how these phases can be switched by tuning $B$, $n$ and $D$, respectively. By flipping $B$, both spin and valley flavors are flipped, leading to the reversion of orbital magnetization. By flipping $n$, the counter part on top and bottom layers (of same valley but different spin) are flipped, while leaving the other two unchanged, which also agrees with our observation on electrically switchable Chern insulators (Fig.3). As for the observed hysteresis on $D$, we attribute it to random selections within the four "ALL" states under a given $D$ (Fig.4), which are not distinguishable in experiments at $n = B = 0$. However, by applying a magnetic field to polarize the system to a specific phase, we have indeed found only one species of hysteresis loops (Fig.S7). Owing to the partial layer polarization, "ALL" phases should be stabilized with an interlayer electric field, namely $|D| = 0.1$ V/nm as we found. At very high or zero electric fields, the phase should vanish again, losing stability against a fully layer polarized insulating (LPI) state or an equally polarized layer antiferromagnetic (LAF) insulating state.

## Supplementary References